\documentstyle[floats, aps, epsfig]{revtex}

\draft

\begin{document}

\twocolumn[\hsize\textwidth\columnwidth\hsize\csname @twocolumnfalse\endcsname

\title{Magnetic Field Pinning of a Dynamic Electron-Spin-Resonance Line
in a GaAs/AlGaAs Heterostructure}

\author{Chris Hillman}
\address{Department of Electrical Engineering, University of California, Los
Angeles, CA 90095-1594}

\author{H.~W.~Jiang}
\address{Department of Physics and Astronomy, University of California at Los
Angeles, Los Angeles, CA 90095}

\date{\today}

\maketitle

\begin{abstract}
Electrically detected electron spin resonance (ESR) is used to study the
hyperfine interaction of the two-dimensional electrons and the nuclei of the
host lattice in a GaAs/AlGaAs heterostructure. Under the microwave and radio-
frequency double excitations, we have observed that the ESR line can be pinned
in a very narrow range of magnetic field - in the vicinity of the nuclear
magnetic resonance (NMR) of the nuclei of the GaAs crystal.  Our observations
suggest that this pinning effect is the result of a competition process between
the ESR induced dynamic nuclear polarization and the NMR induced depolarization.
\end{abstract}

]

\narrowtext
Electrically detected magnetic resonance in variety of GaAs based devices has demonstrated the strong coupling of the two-dimensional electron gas (2DEG) with
the nuclear spins of its host crystal. \cite
{Dobers,Berg,Kane,Wald,Kronmuller,Vitkalov} During the electron spin relaxation,
the hyperfine interaction triggers a mutual flip of electron and nuclear
spins to dynamically polarize the nuclear spins in the vicinity of the 2DEG. As
a result of this nuclear polarization and the extremely long nuclear spin
relaxation time in the quantum Hall regime, a wealth of dynamic ESR effects,
such as Overhauser shift, hysteresis, and instability, are exhibited.

Recently, there are increasing activities using the electron spins in group III-
V semiconductors for spin based electronics devices, dubbed as spintronics
devices, \cite {Spintronics} and for quantum information processing. \cite
{Vrijen}  Although III-V semiconductors are very attractive candidates for these
applications due their potential for electron-spin to photon-polarization
transfer, all the nuclear isotopes in the crystals have non-zero angular
momentum and a strong hyperfine coupling to the electrons.  In order to
precisely manipulate the electron spins under ESR condition, the aforementioned
consequences of the strong hyperfine interaction have to be confronted and
better understood.

In an early experiment, Dobers et al. has shown that the nuclear magnetic moment
can be altered by applying an radio-frequency (RF) radiation at NMR \cite
{Dobers}.  Thus, the application of the RF radiation provides control of the
local nuclear magnetic moment which in turn has a strong influence on the
dynamics of the ESR of the 2DEG.  In this paper, we further explore the effect
of the RF radiation on the dynamic ESR of a 2DEG in a GaAs/AlGaAs
heterostructure by using RF radiation with an extended
range of power and field-sweep rate.  We report an experimental result that the
dynamic ESR signal can be pinned in a narrow range of magnetic field in the
presence of RF radiation.

The sample used in this experiment is a modulation doped
GaAs/Al$_{0.3}$Ga$_{0.7}$As heterostructure fabricated by molecular beam
epitaxy.  The details of the sample structure are described elsewhere \cite
{Jiang}.  The mobility of the sample is 800,000 $cm^{2}/V-sec$ with a density of
about $1.6\times~10^{11}\text{~cm}^{-2}$, which vary slightly for different
cool-downs.  Photolithography techniques pattern a 100 $\mu$m by 100 $\mu$m van
der Pauw mesa structure with Ohmic contacts in its four corners.
The sample is immersed in the bath of the pumped liquid helium at a temperature
of 1.3 K.  Applied normal to the plane of a 2DEG is a dc magnetic field produced
by a superconducting magnet.  Microwave fields transverse to the sample are
generated using a short microstrip line that is machined on a commercially
available microwave laminate board. Transition from a coaxial cable to the
microstrip line is made with an SMA to microstrip board end launch adapter.
Opposite the launcher, the microstrip is
terminated with a short to ground. The microwave magnetic field circulates about
the strip so that when the sample is placed flat on the surface of the
microstrip, the ac field is parallel to the 2DEG. Additionally, a single turn
coil is placed around the sample, through which an RF field is superimposed onto
the microwave field.  Use of an RF combiner allows us to supply up to three RF
frequencies to the coil and sample simultaneously which enables single magnetic
field NMR excitation of the Ga$^{69}$, Ga$^{71}$, and As$^{75}$ isotopes.

The experiment is carried out in the quantum Hall effect regime near the Landau
level filling factor $\nu$, given by hn/cB, of 3. The odd filling factor regime
provides an ideal testing ground to study the magnetic resonance
\cite{Dobers,Stein,Jiang}.  In this regime, the electrons of the 2DEG at the
Fermi level are spin polarized and the resistivity is very sensitive to the spin
population and the electronic temperature. Under microwave radiation ($>$ 1 mW
at the source) the ESR can be detected directly in R$_{xx}$, without microwave
modulation, as shown in Figure 1. In literature, it is suggested that the change
of the conductivity during ESR is due to absorptive heating of the 2DEG \cite
{Meisels}.

\begin{figure}[!t]
\begin{center}
\epsfig{file = 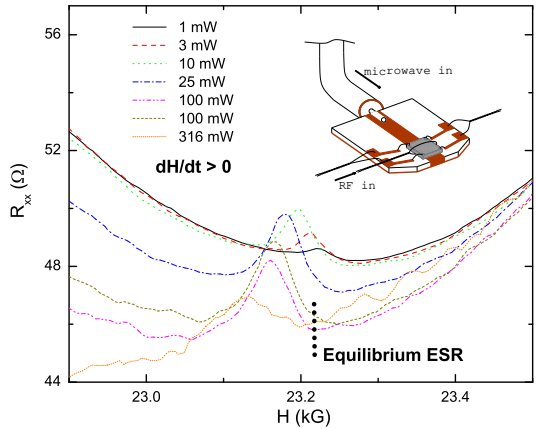, width = 7cm, clip=}
\end{center}
\caption{Typical traces of resistively detected ESR spectra for
different microwave powers around $\nu$=3.  Note the ESR peak
shifts to lower B for an increasing power showing strong dynamic
polarization and line-broadening.  Inset: schematic of the
microwave and radio frequency sample coupling
structure. The microwave source power was 100 mW.}
\label{Figure 1}
\end{figure}

Here the linewidth of the observed ESR is not the intrinsic line width, but
instead depends primarily on the sweep-rate and sweep direction.  Additionally,
we find that the ESR peak position dependents not only on these sweep parameters
but also on the microwave power. Shown in Figure 1 are a set of rather broad ESR
lines, about 100 G, which show that as the power of the microwave is increased,
the ESR peak shifts towards lower B, further away from its thermal equilibrium
position, indicated by the dashed line.

These effects are due to dynamic polarization \cite{Jeffries,Dobers} of nuclei
in the vicinity of the 2DEG.  As the nuclear spins become polarized, local
electrons experience an effective magnetic field in the direction opposite to
the applied field.  For a fixed excitation frequency, the ESR moves to lower
applied magnetic fields for increasing nuclear polarization. This is commonly
known as the Overhauser effect. So we believe Figure 1 indicates increasing
nuclear
polarization for larger microwave
power.  Between each of these traces the microwave power is turned-off
and the sample is irradiated with three RF frequencies corresponding to the NMR
of the three nuclear isotopes of the lattice at a single magnetic field.  The
magnetic field is then slowly swept down through the NMR field. We have found
that doing so apparently returns the nuclear polarization to near equilibrium
and helps to ensure repeatability between measurements.  It is important to
mention here that these traces were
obtained by rapid increasing-field scans ($>$ 30 G/Sec) while irradiating the
sample with the same
microwave frequency. A fast up-field scan, in which the dB/dt and the rate of
change of the resonance position dB$_{res}$/dt are of opposite sign, can
always cross to produce an ESR peak. In contrast, with sufficiently strong
microwave power, a slow down-field scan can result in a locking of ESR-position
to the external applied magnetic field and no ESR peak can be observed. \cite
{Dobers}  It is apparent from the discussion that the ESR line moves dynamically
under strong microwave radiation.  Its position is not only power dependent but
also can evolve as a function of time.

\begin{figure}[!b]
\begin{center}
\epsfig{file = 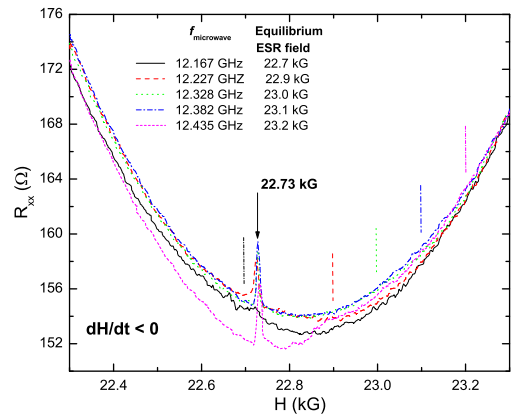, width = 7.5cm, clip=}
\end{center}
\caption{Resistivity $\rho_{xx}$ as a function of the magnetic
field under microwave and RF double excitations.  The ESR signal
is observed as a narrow line ($<$ 10 G) for different microwave
excitation frequencies.  The line position is largely independent
of the microwave frequency. The RF excitation was 16.503 MHz
at 3 mW source power.}
\label{Figure 2}
\end{figure}

As indicated in the previous paragraph, the addition RF radiation dramatically
changes the behavior of the ESR line. Now we would like to present the main
result of the paper: the effect of the RF radiation on the dynamic ESR.  Shown
in Figure 2, the ESR signal, under additional RF irradiation, is exhibited as a
significantly narrower line that is observed in a very narrow range of magnetic
field, dependent primarily on the radio frequency.  The narrow line can be
described as approximately Lorentzian
having a half-width at full maximum of less than 10 G and as small as 2 G.  The
traces shown in Figure 2 are produced by slow decreasing-field scans (dB/dt = -5
G/Sec) while applying a microwave and RF excitation simultaneously.  The
observed ESR is largely independent of the applied microwave excitation
frequency and the lines are effectively pinned to a field determined by the
radio frequency. Looking at the figure, varying the microwave
frequency from 12.435 GHz to 12.167 GHz,
the peak position is almost at the exact same field in each trace. The equilibrium ESR field
values for the  excitation frequencies used are indicated as the dashed lines in
the figure. We found that the narrow line is detectable for fields as far as 4
kG from the equilibrium ESR field.

\begin{figure}[!b]
\begin{center}
\epsfig{file = 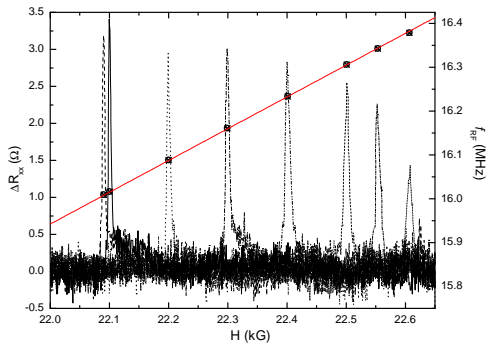, width = 7.5cm, clip=}
\end{center}
\caption{The swept magnetic field ESR spectra at a sequence of RF frequencies.
The non-resonant background of the signal is subtracted for
clarity. On the same figure, the RF frequency (left axis) isplotted as a
function of the peak position.  The slope of the line
confirms that the peak is in the vicinity of the NMR field of
As$^{75}$ nucleus.  The microwave excitation was 12.375 GHz at while both microwave
and RF source powers were set at 100 mW.}
\label{Figure 3}
\end{figure}

The narrow peak occurs approximately at the field corresponding to the NMR
frequency of
$^{75}$As. Figure 3 shows the position of the line as a function of the RF
frequency.
The plot of frequency versus peak position is displayed on the same figure. The
slope of the fitted straight line in the graph, 0.79 MHz/T, confirms the ESR
signal is in the vicinity of the NMR of the As$^{75}$ nucleus.  Similar
experiments, not shown, have been done in the RF frequency range for the
Ga$^{69}$ and Ga$^{71}$ nuclei, where the same conclusion can be reached as that
for the As$^{75}$ case. This effect can be observed for an RF field as small as
$\sim$ 10$^{-7}$ T (the field is calculated based on input power, cable
attenuation and RF coil inductance).  We found that the narrow
peak is exhibited very weakly or not at all when the equilibrium ESR field is
less than NMR magnetic field value.  We also found that the field position of
the
narrow ESR signal depended slightly on the applied power of the microwave
radiation.
For example, a shift about 10 G is observed when the source microwave power is
varied
from 1 mW to 32 mW.

\begin{figure}[!b]
\begin{center}
\epsfig{file = 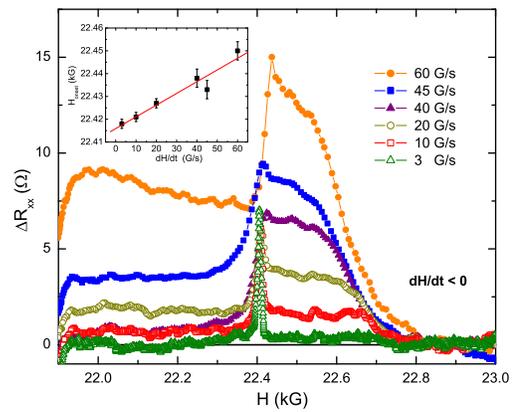, width = 7.5cm, clip=}
\end{center}
\caption{The ESR spectra for a sequence of field sweep rates. As the sweep rate
approaches zero, the ESR signal is exhibited in a narrow range of field, similar
to the traces of Figure 2.  However, for fast scan traces, a broad signals is
observed prior to the onset of the narrowed peak.  The field for this onset is plotted against the sweep rate in the inset. The measurements were conducted with 16.285 MHz and 12.139 GHz at 10 mW and 32 mW source powers respectively.}
\label{Figure 4}
\end{figure}

The last measurement described here demonstrates the field sweep rate dependence
of the observed ESR signal.  Figure 4 shows the evolution of the ESR line for
sweep rates ranging from - 60 G/s to - 3 G/s (the baseline is subtracted for
clarity). For fast scans a broad ESR signal can be seen to rise
rapidly from near the equilibrium ESR field of about 22.7 kG. However, as the
sweep rate is reduced, the broad signal
reduces in intensity leaving the narrow line near the NMR field as the prominent
feature, reproducing the signal shown in Figure 2.  We would like to note that
the
observation reported earlier \cite{Dobers} shows close resemblance to our fast
sweep rate traces.  Also note in the figure that the peak position appears to
shift toward
higher field for increasing field sweep rates. We believe the scanning rate
dependence reveals the competition between the ESR induced dynamic nuclear
polarization and the NMR induced
nuclear depolarization, as will be discussed later.

Having established the experimental fact that the exhibition of a narrow ESR
line can be pinned near the field corresponding to NMR condition, we would now
like to discuss the possible origin of this effect.

In magnetic resonance literature, a common case of NMR detected ESR is known to
be spectrum hole burning \cite {Poole,Slichter}. An inhomogeneously broadened
ESR line can be considered as a superposition of independent ``spin packets" in
different local nuclear environments.  Traditionally, a saturated ESR may be de-
saturated by applying an NMR excitation at a field within the broadened ESR.
The NMR excitation modifies the nuclear field experienced by the electrons
changing the ESR condition at that instantaneous magnetic field.  First of all,
the hole burning picture should only apply to electron spins that are localized
(ex. donors in semiconductor)where the statistical variance of the nuclear spin
orientation gives rise the inhomogeneous broadening.  In our case, the extend of
the localized wave function at $\nu$ =3 is much too large ($>>$ 20 nm) to
consider the 2DEG as localized electrons for the purpose of hole burning, as
many wave functions overlap at a given site of nucleus.  The inhomogeneous line
should thus be averaged out. Furthermore, no saturation is ever observed in our
experiment due to the relatively short spin relaxation times of the 2DEG.

One can also imaging that the sweeping of fielded through the NMR would produce
an adiabatic fast
passage of the nuclear spin. When one sweeps through the NMR fast enough in
comparison to the spin relaxation time T$_{1n}$ of the nuclei, the sign of the
magnetization is reversed\cite{Slichter}.  Indeed, the T$_{1n}$ in our
experiment is
determined to be $\sim$250 Sec by the time dependence of the Overhauser shift
which is
much longer than the field scan passage time.  However, we found the ESR
position, after scan through the NMR, is still left behind in the low-
field side of the NMR, showing no signature of nuclear polarization sign
reversal.

We would like to suggest a competition process which we found to be consistent
with our observations.  This process is a consequence of the competition between
the ESR induced dynamic nuclear polarization and the NMR induced nuclear
depolarization.  The 2DEG at the Fermi level are well polarized along the
applied field direction.  When microwave radiation is applied at ESR,
electron spins are flipped.  As they relax, angular momentum is conserved by
mutual nuclear spin flips, through the scalar hyperfine interaction.  Due to the very long nuclear relaxation time T$_{1n}$, the nuclear spin polarization steadily
increases while the contiguously excited electron spins relax within their
relatively short, T$_{1e}$.  A slow sweep of the field produce a steady polarization
rate which pushes the ESR continuously towards lower B.  However,
once the NMR condition is satisfied, depolarization of nuclear spin system
starts as the RF field tends to equalize the nuclear spin populations.  This
depolarization process, in contrast, would restore the ESR to its equilibrium
field. As the two process compete, the ESR line is ``squeezed" into a field
range.  Since the linewidth of the NMR is extremely
narrow in comparison to ESR line, an equilibrium would be established always in
the vicinity of the NMR field.  The exact position would, however, dependents on
the relative amplitude of the microwave to the RF field.  Using a rough model of
rate equation, we have performed numerical calculations to simulate this effect.
To describe the dynamic nuclear polarization, the rate equation identical to
that used in equation 4 of the reference \cite {Seck} has been used.  The
standard rate equation for the
NMR-induced depolarization is also used\cite{Abragam}.   The electron
relaxation times T$_{1e}$ and T$_{2e}$, used in reference \cite {Seck} for n-
GaAs system were also applied along with the experimentally obtained T$_{1n}$.
By combining these rate equations and solving them numerically, we were be able
to produce a sharp ESR line near the NMR field.  Despite the fact that the
relaxation rates are not precisely known, we found that the narrow line can be
observed for a broad range of parameters.  Further experimental
evidence supporting the proposed mechanism can be found in Figure 4. As the
sweep rate, or the absorption rate, increases (indicated by signal strength),
the difference between the instantaneous field and the dynamic center field of
the ESR diminishes. Therefore, smaller NMR absorption/depolarization, which can
be obtained readily in the high-field tail region of the NMR spectrum, is required to overtake the dynamic polarization.  As a consequence, the onset of the narrow ESR signal occurs at higher magnetic field as shown in the inset of Figure 4.  The evolution of the ESR signal for from slow to fast scan, as observed in Figure 4,
is in fact fairly well reproduced in our numerical simulations.

In the large collection of magnetic resonance literature, the effect reported
here does not appear to be a common effect. In fact, we are not aware of any
similar experimental report.  We believe the main reason for observing this
effect in our system is because the very strong hyperfine coupling and fast
dynamic nuclear polarization rate in this system.  A large Overhauser shift, of
order of 1 kG, is not common in other conduction electron spin systems.
Furthermore, the slow nuclear spin relaxation T$_{1n}$, due to the
absence of the density of states in the quantum Hall regime, makes the dynamic
nuclear polarization to be even more prominent. \cite{Berg,Vagner} In light of
spintronics and quantum information processing applications, we envision that
the effect described in this paper can be used potentially for locking a dynamic
ESR line to a very narrow range of fields for the spin manipulations.

We would like to thank E. Yablonovitch, K. Holczer and S. Brown for stimulating
discussions.  The work is sponsored by the Defense Advanced Research Project
Agency under grant number DAAD19-00-1-0172.

\end{document}